\documentclass[10pt,twocolumn,letterpaper]{article}

\usepackage{iccv}
\usepackage{times}
\usepackage{epsfig}
\usepackage{graphicx}
\usepackage{amsmath}
\usepackage{amssymb}
\usepackage{mathrsfs}
\usepackage{amsthm,amsmath,amssymb}
\usepackage{pdfpages}
\usepackage{tabularx}
\usepackage[T1]{fontenc}
\usepackage[utf8]{inputenc}
\usepackage{authblk}
\usepackage{eso-pic}

\usepackage[breaklinks=true,bookmarks=false]{hyperref}

\iccvfinalcopy 

\def\iccvPaperID{****} 

\ificcvfinal\fi

\begin{document}

\title{Overfitting the Data: Compact Neural Video Delivery via Content-aware Feature Modulation}

\author[1\thanks{Equal Contribution.}]{Jiaming Liu}
\author[2*\thanks{This work was done when Jiaming Liu was an intern at Intel Labs China supervised by Ming Lu}]{Ming Lu}
\author[1]{Kaixin Chen}
\author[1]{Xiaoqi Li}
\author[1]{Shizun Wang}
\author[1]{Zhaoqing Wang}
\author[3]{Enhua Wu}
\author[2]{Yurong Chen}
\author[1\thanks{Chuang Zhang is responsible for correspondence.}]{Chuang Zhang}
\author[1]{Ming Wu}

\affil[1]{Beijing University of Posts and Telecommunications}
\affil[2]{Intel Labs China}
\affil[3]{State Key Lab of Computer Science, IOS, CAS $\&$ FST, University of Macau}
\affil[ ]{\tt\small {\{liujiaming,zhangchuang\}@bupt.edu.cn},lu199192@gmail.com}



\maketitle
\ificcvfinal\thispagestyle{empty}\fi

\begin{abstract}
Internet video delivery has undergone a tremendous explosion of growth over the past few years. However, the quality of video delivery system greatly depends on the Internet bandwidth. Deep Neural Networks (DNNs) are utilized to improve the quality of video delivery recently. These methods divide a video into chunks, and stream LR video chunks and corresponding content-aware models to the client. The client runs the inference of models to super-resolve the LR chunks. Consequently, a large number of models are streamed in order to deliver a video. In this paper, we first carefully study the relation between models of different chunks, then we tactfully design a joint training framework along with the Content-aware Feature Modulation (CaFM) layer to compress these models for neural video delivery. {\bf With our method, each video chunk only requires less than $1\% $ of original parameters to be streamed, achieving even better SR performance.} We conduct extensive experiments across various SR backbones, video time length, and scaling factors to demonstrate the advantages of our method. Besides, our method can be also viewed as a new approach of video coding. Our primary experiments achieve better video quality compared with the commercial H.264 and H.265 standard under the same storage cost, showing the great potential of the proposed method. Code is available at:\url{https://github.com/Neural-video-delivery/CaFM-Pytorch-ICCV2021}
\end{abstract}

\section{Introduction}

Internet video is achieving explosive growth over the past few years, which brings a huge burden to the video delivery infrastructure. The quality of video heavily depends on the bandwidth between servers and clients. Techniques at both sides evolve over time to handle the scalability challenges at Internet scale. Inspired by the increasing computational power of client/server and recent advances in deep learning, several works are proposed to apply Deep Neural Networks (DNNs) to video delivery system \cite{kim2020neural,yeo2018neural}. The core idea of these works is to stream both the low resolution video and content-aware models from servers to clients. The clients run the inference of models to super-resolve the LR videos. In this manner, better user Quality of Experience (QoE) can be obtained under limited Internet bandwidth.

In contrast to current approaches on Single Image Super-Resolution (SISR) \cite{shi2016real,dong2014learning,lim2017enhanced,zhang2018image,kim2016accurate} and Video Super-Resolution (VSR) \cite{caballero2017real,wang2019edvr,chan2020basicvsr}, content-aware DNNs leverage neural network's overfitting property and use the training accuracy to achieve high performance. Specifically, a video is first divided into several chunks, and then a separate DNN is trained for each chunk. The low resolution chunks and corresponding trained models are delivered to the clients over the Internet. Different backbones \cite{shi2016real,dong2014learning,lim2017enhanced,zhang2018image,kim2016accurate} can be used as the DNN for each chunk. This kind of DNN-based video delivery system has achieved better performance compared with commercial video delivery techniques like WebRTC \cite{kim2020neural}.  

Although it is promising to apply DNNs to video delivery, existing methods still have several limitations \cite{lee2020neural}. One major limitation is that they need to train one DNN for each chunk, resulting in a large number of separate models for a long video. This brings additional storage and bandwidth cost for the practical video delivery system. In this paper, we first carefully study the relation between models of different chunks. Although these models are trained to overfit different chunks, we observe the relation between their feature maps is linear and can be modelled by a Content-aware Feature Modulation (CaFM) layer. This motivates us to design a method, which allows the models to share the most of parameters and preserve only private CaFM layers for each chunk. \cite{he2019modulating} proposed a closely related method for continual modulation of restoration levels. While they aimed to handle arbitrary restoration levels between a start and an end level, our goal is to compress the models of different chunks for video delivery. The reader is encouraged to review their work for more details. However, directly finetuning the private parameter fails to obtain competitive performance compared with separately trained models. Therefore, we further design a tactful joint training framework, which trains the shared parameters and private parameters simultaneously for all chunks. In this way, our method can achieve relative better performance compared with individually trained models.

Apart from video delivery, our method can also be considered as a new approach of video coding. We conduct primary experiments to compare the proposed approach against commercial H.264 and H.265 standards under the same storage cost. Our method can achieve higher PSNR performance thanks to the overfitting property, showing the great potential of the proposed approach.

Our contributions can be concluded as follows:

\begin{itemize}
\item We propose a novel joint training framework along with the content-aware feature modulation layer for neural video delivery.

\item We conduct extensive experiments across various SR backbones, video time length, and scaling factors to demonstrate the advantages of our method.

\item We compare with commercial H.264 and H.265 standard under the same storage cost and show promising results thanks to the overfitting property. 
\end{itemize}

\begin{figure*}[ht]
\begin{center}
\includegraphics[width=15cm, height=4.7cm]{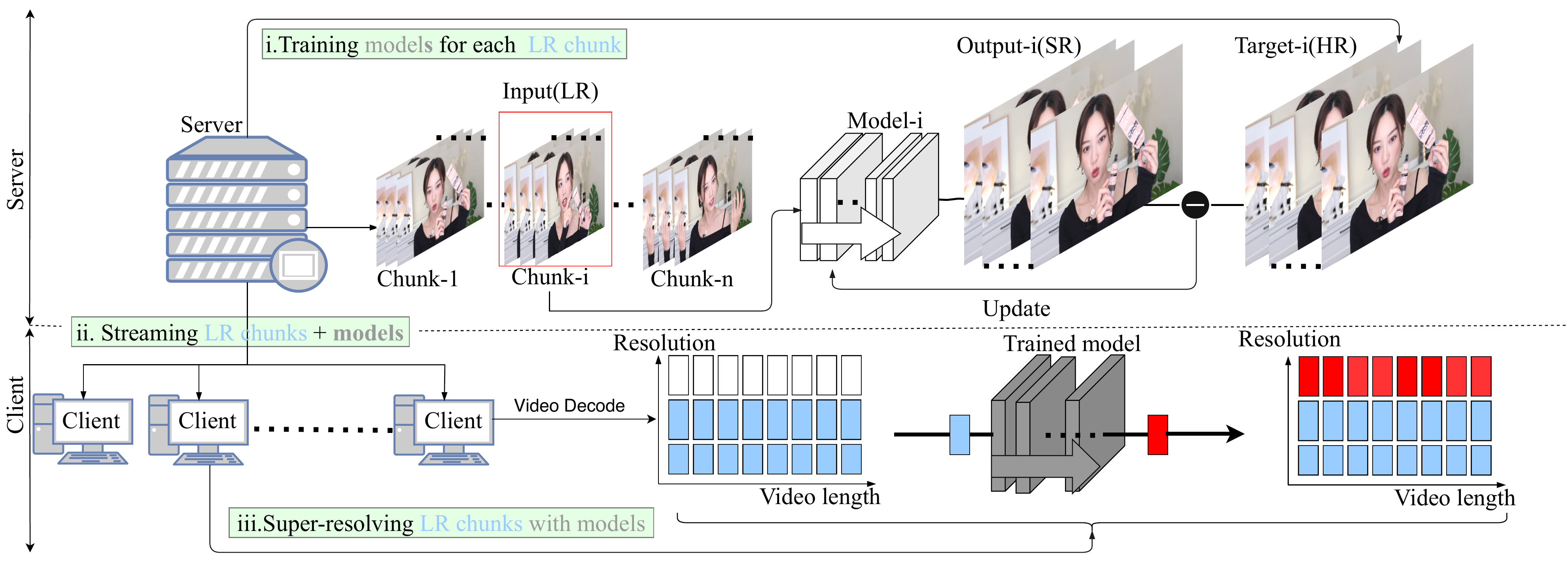}
\end{center}
   \caption{The whole procedure of adopting content-aware DNNs for video delivery. A video is first divided into several chunks and the server trains one model for each chunk. Then the server delivers LR video chunks and models to client. The client runs the inference to super-resolve the LR chunks and obtain the SR video.}
\label{fig:nvd}
\end{figure*}

\section{Related Work}

{\bf DNN-based Image Super-Resolution} SRCNN \cite{dong2014learning} is the pioneering work that introduces DNN to the SR task. Their DNN consists of three stages, namely feature extraction, non-linear mapping, and image reconstruction. With the development of DNN, following the pipeline of SRCNN, plenty of DNN-based methods are proposed to improve the performance of SISR. For example, VDSR \cite{kim2016accurate} uses a very deep DNN to learn the image residual instead of HR image. Inspired by ResNet \cite{he2016deep}, SRResNet \cite{ledig2017photo} introduces Residual Block to SR and improves the capability of DNN. EDSR \cite{lim2017enhanced} modifies the structure of SRResNet and removes the Batch Normalization layer \cite{ioffe2015batch}, further advancing the SR results. RDN \cite{zhang2018residual} proposes to adopt dense connection \cite{huang2017densely} in order to fully use previous layers' information. RCAN \cite{zhang2018image} studies the attention mechanism and presents deeper DNs for SR. However, RCAN is computationally complicated, which limits its practical usage. To reduce the computational cost, many works are proposed for efficient SR. ESPCN \cite{shi2016real} uses LR image as input and upsamples the feature map by the PixelShuffle layer at the end of DNN. LapSRN \cite{lai2018fast} proposes the Laplacian Pyramid network for fast and accurate SR, which progressively reconstructs the sub-band residuals of HR images. FALSR \cite{chu2019fast} uses Neural Architecture Search (NAS) to search the lightweight and accurate networks for SR. LAPAR \cite{li2020lapar} presents a method based on linearly-assembled pixel-adaptive regression network, which learns the pixel-wise filter kernel for SR. All of those methods are external methods, which train the model on large-scale image databases like DIV2K \cite{Agustsson_2017_CVPR_Workshops} and test on certain benchmark databases. However, external methods fail to explore the overfitting property of DNNs, which are useful for practical video delivery systems. 

{\bf DNN-based Video Super-Resolution} In contrast to image super-resolution, video super-resolution can additionally utilize temporal information for SR. Therefore, temporal alignment plays an important role and should be extensively studied. VESPCN \cite{caballero2017real} estimates the motions between neighboring frames and performs image warping before feeding neighboring frames into SR network. However, a precise estimation of optical flow is often not tractable. TOFlow \cite{xue2019video} proposes a task-oriented flow tailored for specific video processing tasks. They jointly train the motion estimation component and video processing component in a self-supervised manner. DUF \cite{jo2018deep} avoids the accurate explicit motion compensation by training a network to generate dynamic upsampling filters and a residual image. EDVR \cite{wang2019edvr} improves the performance of VSR from two aspects. First, they devise an alignment module to handle large motions, in which frame alignment is performed at the feature level using deformable convolutions in a coarse-to-fine manner. Second, they apply attention mechanism both temporally and spatially, aiming to emphasize important features for subsequent restoration. In order to reduce the computational cost of VSR, FRVSR \cite{sajjadi2018frame} presents a recurrent framework that uses the previous SR to super-resolve the subsequent frame. Their recurrent frames naturally encourage temprally consistency and reduce the computational cost by warping only one image in each step. TecoGAN \cite{chu2020learning} explores the temporal self-supervision for GAN-based VSR. They propose a temporal adversarial learning method that achieves temporally coherent solutions without sacrificing spatial detail. BasicVSR \cite{chan2020basicvsr} reconsider some most essential components for VSR by reusing some existing components added with minimal redesigns. They achieve appealing improvements in terms of speed and restoration quality in comparison to state-of-the-art algorithms. All these VSR approaches also belong to external methods that fail to explore the overfitting property of DNN. Apart from this, handling temporal alignment brings huge additional computational and storage cost, which limits their practical applications in resource-limited devices like mobile phone.

{\bf Neural Video Delivery} NAS \cite{yeo2018neural} is a new and promising practical Internet video delivery framework that integrates DNN based quality enhancement. It can solve the video quality degradation problem when Internet bandwidth becomes congested. NAS can enhance the average Quality of Experience (QoE) by $43.08\%$ using the same bandwidth budget, or saving $17.13\%$ of bandwidth while providing the same user QoE. The core idea is to leverage DNN's overfitting property and use the training accuracy to deliver high enhancement performance. Many following works are proposed to apply the idea of NAS to different applications, like UAV video streaming \cite{xiao2019sensor}, Live Streaming \cite{kim2020neural}, 360 Video Streaming \cite{dasari2020streaming,chen2020sr360}, Volumetric Video Streaming \cite{zhang2020mobile}, and Mobile Video Streaming \cite{yeo2020nemo}, etc. In this paper, the proposed method can further reduce the bandwidth budget by sharing most of the parameters over video chunks. Therefore, only a small portion of private parameters are streamed for each video chunk.

\section{Proposed Method}
Neural video delivery \cite{kim2020neural,yeo2018neural} aims at utilizing DNNs to save bandwidth when delivering Internet videos. Different from traditional video delivery systems, they replace HR video with LR video and content-aware models. As shown in Fig. \ref{fig:nvd}, the whole procedure includes three stages: (i) training models for video chunks on the server; (ii) delivering LR chunks along with content-aware models from server to client; (iii) super-resolving LR chunks on client. However, this procedure requires to stream one model for each chunk, resulting in additional bandwidth cost. In this part, we first analyze the relation between models of video chunks in Section \ref{sec:motivation}. Then we introduce a Content-aware Feature Modulation (CaFM) layer to model this relation in section \ref{sec:cafm}. However, fine-tuning the CaFM module on each chunk fails to achieve competitive results. Therefore, we propose a joint training framework in Section \ref{sec:joint}.

\subsection{Motivation}
\label{sec:motivation}

\begin{figure*}[ht]
\begin{center}
\includegraphics[width=16cm, height=4.0cm]{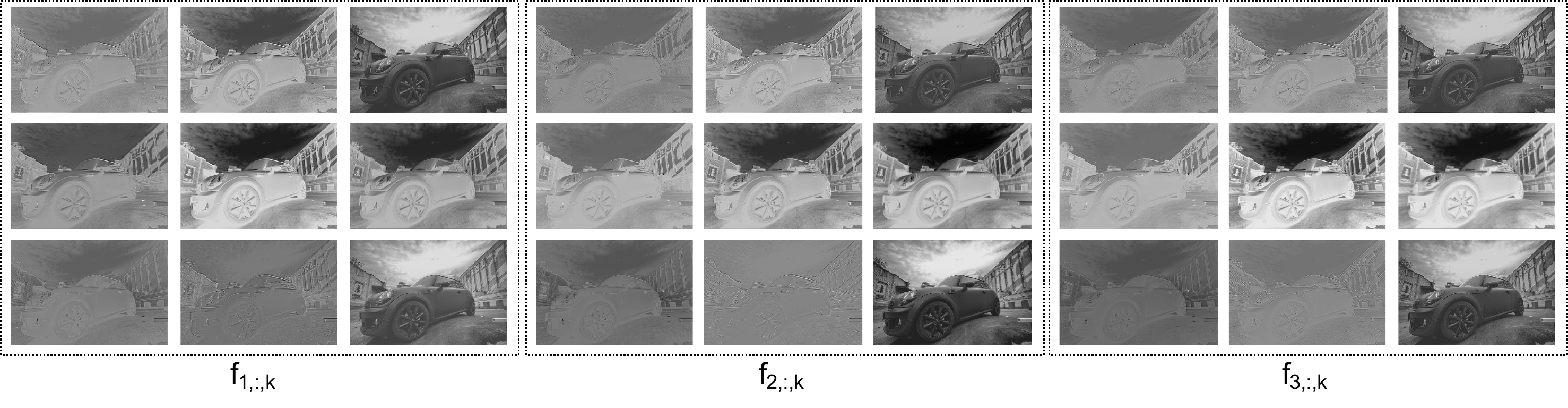}
\end{center}
   \caption{The feature map visualization of different models $S_{1-n}$.}
\label{fig:feature}
\end{figure*}

\begin{figure}[ht]
\begin{center}
\includegraphics[width=7cm, height=9cm]{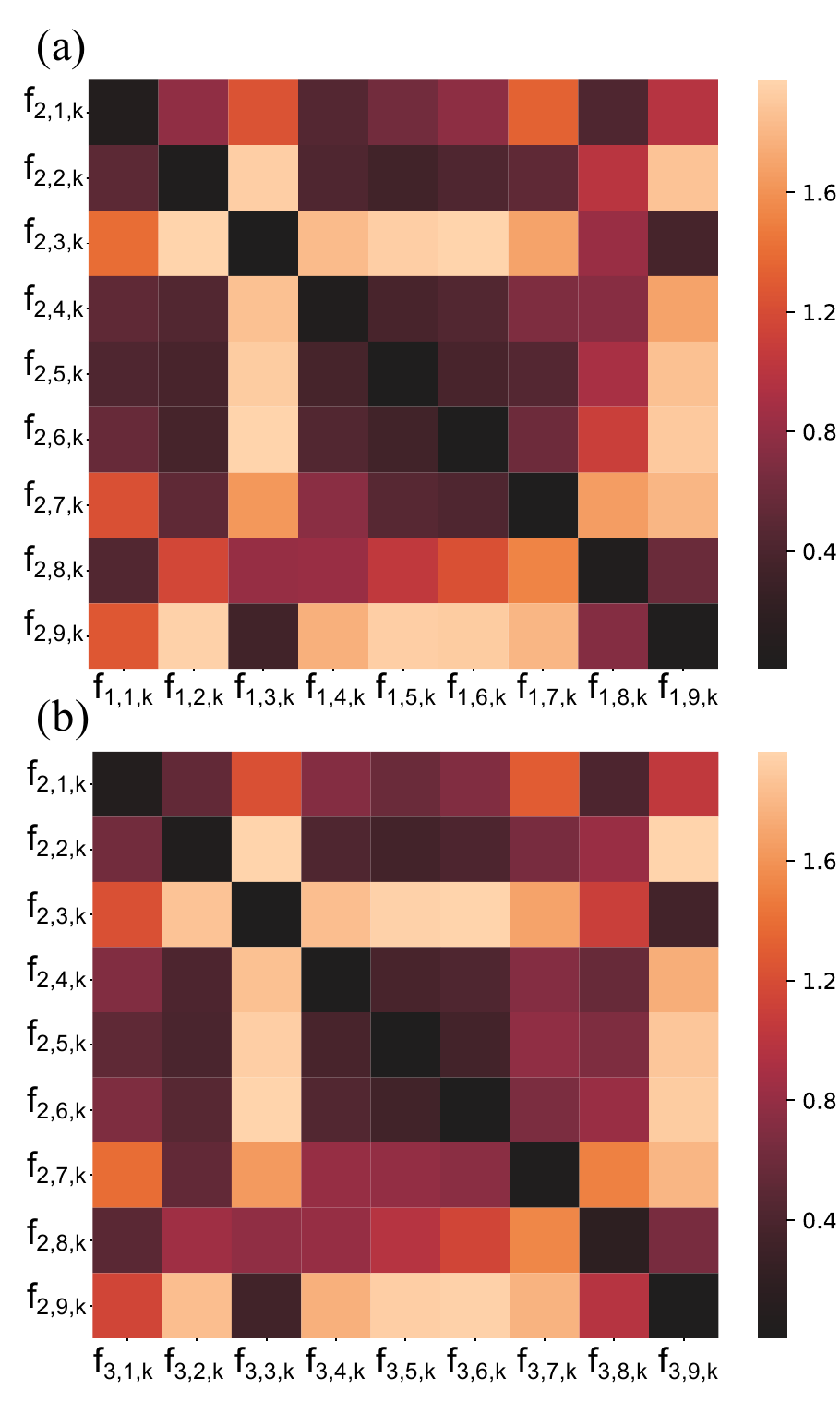}
\end{center}
   \caption{The cosine distance matrices of ${f_{1,:,k}}$, ${f_{2,:,k}}$ and ${f_{3,:,k}}$ in Fig. \ref{fig:feature}}
\label{fig:cosine}
\end{figure}

In this part, we adopt the popular architecture EDSR \cite{lim2017enhanced} and analyze the models trained on video chunks. Following former works \cite{yeo2018neural,kim2020neural}, we split a video into $n$ chunks and train $n$ EDSR models $S_1, S_2...S_n$ for these chunks accordingly. Then we empirically analyze $S_1, S_2...S_n$ by feeding a randomly selected image of DIV2K \cite{Agustsson_2017_CVPR_Workshops}. We visualize the feature maps of these $n$ EDSR models in Fig. \ref{fig:feature}. Each image represents the feature map of a certain channel and we only visualize one layer of EDSR for simplicity. Specifically, we denote the feature map as ${f_{i,j,k}} \in {R^{H \times W}}$, where $i$ means the $i^{th}$ model, j means $j^{th}$ channel, and $k$ means the $k^{th}$ layer of EDSR. For the randomly selected image, we can calculate the cosine distance between ${f_{{i_1},{j_1},{k}}}$ and ${f_{{i_2},{j_2},{k}}}$, which measure the similarity between these two feature maps. For the feature maps in Fig. \ref{fig:feature}, we calculate the cosine distance matrix among ${f_{1,:,k}}$, ${f_{2,:,k}}$ and ${f_{3,:,k}}$. As shown in Fig. \ref{fig:cosine}, we observe that although $S_1, S_2...S_n$ are trained on different chunks, the cosine distance between corresponding channel is very small according to the diagonal values of the matrices in Fig. \ref{fig:cosine}. We calculate the average of cosine distance among $S_1$, $S_2$, and $S_3$ across all layers, and the results are about 0.16 and 0.04 separately. {\bf This indicates that although different models are trained on different chunks, the relation between ${f_{1,j,k}}$ and ${f_{2,j,k}}$ can be approximately modelled by a linear function.} The aforementioned observation motivates us to share most of the parameters for $S_1, S_2...S_n$ and privatize each DNN with Content-aware Feature Modulation (CaFM), which will be  elaborated in the next part.

\subsection{Content-aware Feature Modulation}
\label{sec:cafm}

\begin{figure*}[ht]
\begin{center}
\includegraphics[width=16cm, height=5cm]{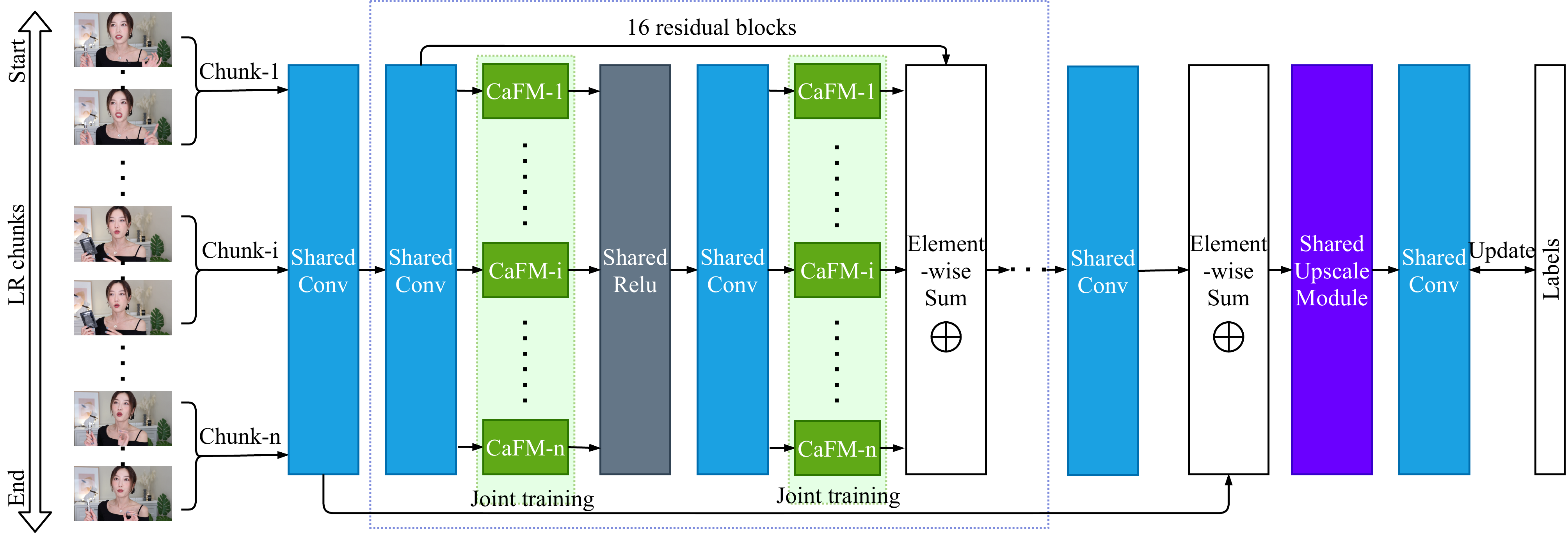}
\end{center}
   \caption{The joint training framework along with CaFM.}
\label{fig:joint}
\end{figure*} 

In this part, we introduce the Content-aware Feature Modulation (CaFM) layer into baseline models to privatize the DNN for each video chunk. The overall framework is shown in Fig. \ref{fig:joint}. As mentioned in the motivation, CaFM aims to manipulate the feature map and adapt the model to different chunks. Therefore, models of different chunks can share most of the parameters. We represent CaFM as a channel-wise linear function:
\begin{equation}
    CaFM(x_j)=a_j*x_j+b_j, 0<j\leq C
\end{equation}
where $x_j$ is the $j^{th}$ input feature map, $C$ is the number of feature channels, $a_j$ and $b_j$ are the channel-wise scaling and bias parameters respectively. We add CaFM to modulate the output feature of each filter for the baseline models. Taking EDSR as the example, the parameter of CaFM takes up about {\bf 0.6\%} of EDSR. Therefore, for a video with n chunks, we can reduce the size of models from n EDSR to 1 shared EDSR plus n private CaFM. Thus our method can significantly reduce the band-width and storage cost compared with baseline methods. 
\subsection{Joint training framework}
\label{sec:joint}
As mentioned in the above section, we can privatize the DNN of each chunk with CaFM. However, finetuning the added CaFM module cannot achieve competitive results compared with separately trained models. Therefore, we further propose a joint training framework, which simultaneously trains the models of video chunks. Given a LR image $I_{LR}^{i,s}$, where $i$ denotes the $i^{th}$ video chunk and $s$ denotes the $s^{th}$ sample in this chunk, we can obtain the SR image:

\begin{equation}
I_{SR}^{i,s} = g(I_{LR}^{i,s};{W_s},{W_i})
\end{equation}

where $W_s$ is the shared parameter and $W_i$ is the parameter of CaFM for $i^{th}$ video chunk. For each video chunk, we can define the reconstruction loss:

\begin{equation}
{L_i} = \frac{{\sum\limits_{s = 1}^S {{{\left\| {I_{HR}^{i,s} - I_{SR}^{i,s}} \right\|}_1}} }}{S}
\end{equation}

Therefore, the loss function of joint training can be represented as:

\begin{equation}
    \mathcal{L}=\sum\limits_{i = 1}^n {{L_i}} 
\end{equation}

During training, we uniformly sample the images from video chunks to construct the training data. All images are used to update the shared parameter ${W_s}$, while images of $i^{th}$ video chunk are used to update the corresponding CaFM parameter ${W_i}$.
\section{Experiments}
In this section, we conduct extensive experiments to show the advantages of our method. To evaluate the proposed approach, we construct a Video Streaming Dataset with 4K videos (VSD4K). The details of VSD4K, training protocol, and model structure are given in Section \ref{sec:setup}. In Section \ref{sec:train}, we compare the performance of external learning and content-aware learning based on EDSR \cite{lim2017enhanced} and EDVR \cite{wang2019edvr}. In section \ref{sec:resvsd}, we report the results of our method across different video lengths and scaling factors. We also conduct comprehensive ablation study in Section \ref{sec:abl} to evaluate the contribution of each component. In order to demonstrate the generalization ability of our work, we report results with various popular SR architectures in Section \ref{sec:gen}. Finally, since our method can be also considered as a video coding method, we compare it with the commercial H.264 and H.265 standard in Section \ref{sec:h264}.
\subsection{Experimental Details}
\label{sec:setup}
\textbf{Video Streaming Dataset 4K} 
Public video datasets like Vimeo-90K \cite{xue2019video} and REDS \cite{nah2019ntire} only consist of neighboring frame sequences, which are not suitable for video delivery. Therefore, we collect several 4K videos from YouTube to simulate practical video delivery scenario. We use bicubic interpolation to generate the LR videos following former works \cite{lim2017enhanced,zhang2018image}. We select 6 popular video categories to construct VSD4K. Each category consists of various video lengths including: 15s, 30s, 45s, 1min, 2min, 5min. Detailed information of VSD4K can be found in the supplementary material. 

\textbf{Training details}
Since neural video delivery relies on the overfitting property of DNN, we train and test on the same video. In order to reduce the computational cost, we sample 1 frame out of 10 frames for testing. We use $48 \times 48$ HR patches with corresponding LR patches for training. We adopt Adam optimizer with $\beta_{1}=0.9$, $\beta_{2}=0.999$, $\epsilon=10^{-8}$. L1 loss is adopted as the loss function. The learning rate is set as $10^{-4}$ and decay at different iterations depends on video length. Besides, we set different mini-batch sizes for different video lengths to maintain the computational cost same as the sum of $S_{1-n}$. 


		

\textbf{Model structure} Since our method can be applied to different SR architectures, we conduct extensive experiments with various popular networks (EDSR, VDSR, ESPCN, and SRCNN). We report most of the ablation study results on EDSR for the consideration of simplicity. We add the proposed CaFM module to the output of each convolutional layer in EDSR. The CaFM module is implemented as a $1 \times 1$ depth-wise convolution. Other SR architectures can be modified accordingly.
\subsection{Evaluation on content-aware learning}
\label{sec:train}

\begin{table*}[ht]
\scriptsize
\begin{center}
    

\begin{tabular}{|p{0.14\textwidth}<{\centering}|p{0.06\textwidth}<{\centering}|p{0.08\textwidth}<{\centering}|p{0.03\textwidth}<{\centering}p{0.06\textwidth}<{\centering}p{0.03\textwidth}<{\centering}|p{0.03\textwidth}<{\centering}p{0.06\textwidth}<{\centering}p{0.03\textwidth}<{\centering}|p{0.03\textwidth}<{\centering}p{0.06\textwidth}<{\centering}p{0.03\textwidth}<{\centering}|}
\hline
 & & & &game-45s& & &vlog-45s& & &inter-45s& \\
Method&Model &Dataset & x2 &  x3 &   x4
& x2 &  x3 & x4
&  x2  & x3 & x4
  \\
\hline
 & EDSR-M & DIV2K &36.02&31.20&29.00&46.85&42.70&40.41&38.13&32.00&28.96\\

External learning & EDVR-M&REDS&-&-&28.72&-&-&41.28&-&-&28.90\\
 & EDVR-L&REDS&-&-&29.45&-&-&40.42&-&-&27.30\\
  & EDVR-L&Vimeo-90K&-&-&30.01&-&-&\textcolor{blue}{42.04}&-&-&30.06\\
\hline
Content-aware learning & EDSR-M&VSD4K&\textcolor{blue}{42.11}&\textcolor{blue}{35.75}&\textcolor{blue}{33.33}&\textcolor{blue}{47.98}&\textcolor{blue}{43.58}&41.53&\textcolor{blue}{42.73}&\textcolor{blue}{34.49}&\textcolor{blue}{31.34}\\
\hline
Content-aware learning* & EDSR-M&VSD4K&\textcolor{red}{43.22}&\textcolor{red}{36.72}&\textcolor{red}{34.32}&\textcolor{red}{48.48}&\textcolor{red}{44.12}&\textcolor{red}{42.12}&\textcolor{red}{43.31}&\textcolor{red}{35.80}&\textcolor{red}{32.67}\\
\hline\hline
 & & & &sport-45s& & &dance-45s& & &city-45s& \\
 &  &  & x2 &  x3 &   x4
& x2 &  x3 & x4
&  x2  & x3 & x4
  \\
\hline
 & EDSR-M & DIV2K &46.25&41.23&38.07&40.55&37.72&35.51&37.32&32.20&29.73\\

External learning & EDVR-M&REDS&-&-&40.10&-&-&35.93&-&-&31.52\\
 & EDVR-L&REDS&-&-&40.31&-&-&36.84&-&-&31.91\\
  & EDVR-L&Vimeo-90K&-&-&\textcolor{red}{41.38}&-&-&35.74&-&-&\textcolor{red}{32.28}\\
\hline
Content-aware learning & EDSR-M&VSD4K&\textcolor{blue}{47.81}&\textcolor{blue}{42.28}&39.16&\textcolor{blue}{45.49}&\textcolor{blue}{38.26}&\textcolor{blue}{37.32}&\textcolor{blue}{38.67}&\textcolor{blue}{33.44}&30.86\\
\hline
Content-aware learning* & EDSR-M&VSD4K&\textcolor{red}{48.34}&\textcolor{red}{43.01}&\textcolor{blue}{40.34}&\textcolor{red}{45.71}&\textcolor{red}{38.61}&\textcolor{red}{37.67}&\textcolor{red}{39.87}&\textcolor{red}{34.62}&\textcolor{blue}{31.97}\\
\hline

\end{tabular}
\end{center}
\caption{Comparison of content-aware learning versus external training. EDVR-M, EDVR-L, EDSR-M has 10, 40, 16 resblocks respectively. * denotes training a content-aware DNN for each video chunk. Red and blue indicates the best and the second best results.}
\label{tab:content-aware}
\end{table*}

In this part, we demonstrate the benefit of utilizing DNN's overfitting property for video delivery. Methods like EDSR and EDVR train the DNN on a large-scale dataset and use the trained DNN to super-resolve the LR input. We call this kind of training external learning. However, for neural video delivery, the video is known beforehand. Therefore, we can train the models to overfit each video and achieve better SR performance. This kind of training can be named content-aware learning. Apart from this, a video can be further divided into video chunks and a DNN is trained for each chunk separately. We compare the performance of external learning and content-aware learning in Tab. \ref{tab:content-aware}. As we can see, content-aware learning can achieve much better results compared with external learning. In particular, EDSR with content-aware learning outperforms EDVR with external learning by significant margins. These results prove that content-aware learning is more suitable for video delivery compared with external learning.
\subsection{Evaluation on VSD4K datasets}
\label{sec:resvsd}
\begin{table*}[]
\scriptsize
\begin{center}

\begin{tabular}{|p{0.06\textwidth}<{\centering}|p{0.04\textwidth}<{\centering}p{0.14\textwidth}<{\centering}p{0.03\textwidth}<{\centering}|p{0.04\textwidth}<{\centering}p{0.14\textwidth}<{\centering}p{0.03\textwidth}<{\centering}|p{0.04\textwidth}<{\centering}p{0.14\textwidth}<{\centering}p{0.03\textwidth}<{\centering}|}
\hline
 & &game-15s& & &game-30s& & &game-45s& \\
Scale&x2&x3&x4&x2&x3&x4&x2&x3&x4\\
\hline
M0&42.24&35.88&33.44&41.84&35.54&33.05&42.11&35.75&33.33\\\hline
$S_{1-n}$&42.82&36.42&34.00&43.07&36.73&34.17&43.22&36.72&34.32\\
\hline
Ours&43.13&37.04&34.47&43.08&36.94&34.22&43.32&37.19&34.61\\\hline
Margin&0.31&0.62&0.47&0.01&0.21&0.05&0.10&0.47&0.29\\\hline\hline
 & &vlog-15s& & &vlog-30s& & &vlog-45s& \\
Scale&x2&x3&x4&x2&x3&x4&x2&x3&x4\\
\hline
M0&48.87&44.51&42.58&47.79&43.38&41.24&47.98&43.58&41.53
\\\hline
$S_{1-n}$&49.10&44.80&42.83&48.20&43.68&41.55&48.48&44.12&42.12\\
\hline
Ours&49.30&45.03&43.11&48.35&43.94&41.90&48.45&44.11&42.16
\\\hline
Margin&0.20&0.23&0.28&0.15&0.26&0.35&-0.03&-0.01&0.04\\\hline\hline
 & &inter-15s& & &inter-30s& & &inter-45s& \\
Scale&x2&x3&x4&x2&x3&x4&x2&x3&x4\\
\hline
M0&44.85&37.89&34.94&43.06&35.37&32.30&42.73&34.49&31.34\\\hline
$S_{1-n}$&45.06&38.38&35.47&43.50&36.48&33.42&43.31&35.80&32.67
\\
\hline
Ours&45.35&38.66&35.70&43.65&36.30&33.28&43.37&35.62&32.35
\\\hline
Margin&0.29&0.28&0.23&0.15&-0.18&-0.14&0.06&-0.18&-0.32\\\hline\hline
 & &sport-15s& & &sport-30s& & &sport-45s& \\
Scale&x2&x3&x4&x2&x3&x4&x2&x3&x4\\
\hline
M0&48.20&42.56&39.66&50.36&44.72&41.86&47.81&42.28&39.16\\\hline
$S_{1-n}$&48.43 &43.04&40.38&50.67&45.45&42.94&48.34&43.01&40.34
\\
\hline
Ours&48.48&43.06&40.43&50.74&45.31&42.73&48.24&42.93&40.21
\\\hline
Margin&0.05&0.02&0.05&0.07&-0.14&-0.21&-0.10&-0.08&-0.13\\\hline\hline
 & &dance-15s& & &dance-30s& & &dance-45s& \\
Scale&x2&x3&x4&x2&x3&x4&x2&x3&x4\\
\hline
M0&44.35&37.57&36.18&44.85&37.99&36.67&45.49&38.26&37.32
\\\hline
$S_{1-n}$&44.48&37.69&36.40&44.99&38.13&36.93&45.71&38.61&37.67
\\
\hline
Ours&44.71&37.82&36.61&45.30&39.66&37.22&45.97&39.96&37.84
\\\hline
Margin&0.23&0.13&0.21&0.31&1.53&0.29&0.26&1.35&0.17\\\hline\hline
 & &city-15s& & &city-30s& & &city-45s& \\
Scale&x2&x3&x4&x2&x3&x4&x2&x3&x4\\
\hline
M0&37.89&32.32&29.36&38.90&33.15&30.35&38.67&33.44&30.86
\\\hline
$S_{1-n}$&38.14&32.61&29.67&39.85&34.15&31.30&39.87&34.62&31.97
\\
\hline
Ours&38.17&32.65&29.68&39.70&34.01&31.18&39.75&34.48&31.87\\
\hline
Margin&0.03&0.04&0.01&-0.15&-0.14&-0.12&-0.12&-0.14&-0.10\\
\hline\hline
 & &game-1min& & &game-2min& & &game-5min& \quad\\
Scale&x2&x3&x4&x2&x3&x4&x2&x3&x4\quad\\
\hline
M0&41.82&35.25&32.61&41.89&35.72&33.27&40.62&34.59&32.14\quad\\\hline
$S_{1-n}$&43.24&36.56&33.52&43.20&37.00&34.47&42.47&36.08&33.53\quad\\
\hline
Ours&43.49&37.18&34.33&43.49&37.47&34.80&43.01&36.65&34.07\\\hline
Margin&0.25&0.62&0.81&0.29&0.47&0.33&0.54&0.57&0.54\quad\\\hline

\end{tabular}
\end{center}
\caption{The comprehensive results of our method on VSD4K. We conduct experiments with different video length and scaling factors. We also show the margins between our method and $S_{1-n}$ \cite{yeo2018neural}. }
\label{tab:vsd4k}
\end{table*}
\begin{figure*}[t]
\begin{center}
   \includegraphics[width=14cm,height=11cm]{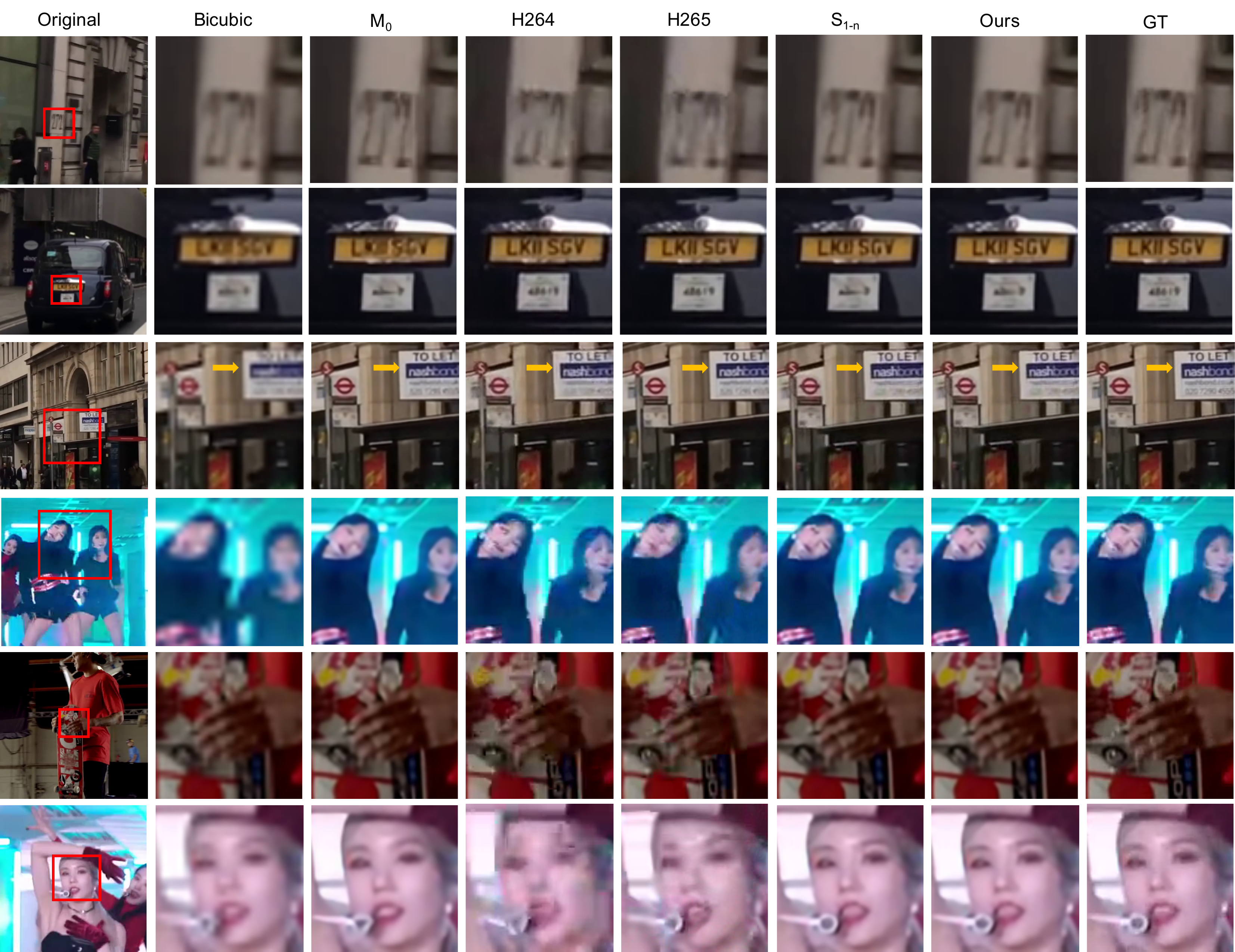}
\end{center}
   \caption{Qualitative comparisons on the VSD4K dataset. Best viewed by zooming x4.}
\label{fig:vsd4k}
\end{figure*}
\begin{table*}[]
\scriptsize
\begin{center}


\begin{tabular}{|p{0.08\textwidth}<{\centering}|p{0.03\textwidth}<{\centering}p{0.14\textwidth}<{\centering}p{0.03\textwidth}<{\centering}|p{0.03\textwidth}<{\centering}p{0.14\textwidth}<{\centering}p{0.03\textwidth}<{\centering}|p{0.03\textwidth}<{\centering}p{0.14\textwidth}<{\centering}p{0.03\textwidth}<{\centering}|}
\hline
 & &game-45s& & &vlog-45s& & &inter-45s& \\\hline
Method&x2&x3&x4&x2&x3&x4&x2&x3&x4\\
\hline
H.264&39.26&36.99&\textcolor{blue}{35.52}&43.45&42.07&41.31&37.99&\textcolor{blue}{36.07}&\textcolor{blue}{35.06}\\\hline
H.265&\textcolor{blue}{39.77}&\textcolor{red}{37.71}&\textcolor{red}{36.42}&\textcolor{blue}{44.24}&\textcolor{blue}{43.09}&\textcolor{red}{42.31}&\textcolor{blue}{38.31}&\textcolor{red}{36.51}&\textcolor{red}{35.58}\\\hline
Ours&\textcolor{red}{43.32}&\textcolor{blue}{37.19}&34.61&\textcolor{red}{48.45}&\textcolor{red}{44.11}&\textcolor{blue}{42.16}&\textcolor{red}{43.37}&35.62&32.35\\\hline
Storage(MB)&19.9&12.6&9.8&19.5&12.5&9.8&19.5&12.4&9.8\\\hline\hline
 & &sport-45s& & &dance-45s& & &city-45s& \\\hline
Method&x2&x3&x4&x2&x3&x4&x2&x3&x4\\
\hline
H.264&40.30&38.09&36.83&31.11&28.32&26.76&36.60&34.18&\textcolor{blue}{32.89}\\\hline
H.265&\textcolor{blue}{41.35}&\textcolor{blue}{39.66}&\textcolor{blue}{38.67}&\textcolor{blue}{32.62}&\textcolor{blue}{30.40}&\textcolor{blue}{29.18}&\textcolor{blue}{37.17}&\textcolor{red}{35.10}&\textcolor{red}{34.03}\\\hline
Ours&\textcolor{red}{48.24}&\textcolor{red}{42.93}&\textcolor{red}{40.21}&\textcolor{red}{45.97}&\textcolor{red}{39.96}&\textcolor{red}{37.84}&\textcolor{red}{39.75}&\textcolor{blue}{34.48}&31.87
\\\hline
Storage(MB)&19.8&12.6&9.9&19.6&12.6&9.8&20.4&12.9&10.1\\\hline

\end{tabular}
\end{center}
\caption{Comparisons with H.264/H.265. Red and blue indicates the best and the second best results.}
\label{tab:h264}
\end{table*}

\begin{table*}[ht]
\scriptsize
\begin{center}


\begin{tabular}{|p{0.04\textwidth}|p{0.03\textwidth}<{\centering}|p{0.03\textwidth}<{\centering}p{0.09\textwidth}<{\centering}p{0.03\textwidth}<{\centering}|p{0.03\textwidth}<{\centering}p{0.09\textwidth}<{\centering}p{0.03\textwidth}<{\centering}|p{0.03\textwidth}<{\centering}p{0.09\textwidth}<{\centering}p{0.03\textwidth}<{\centering}|p{0.03\textwidth}<{\centering}p{0.09\textwidth}<{\centering}p{0.03\textwidth}<{\centering}|}
\hline
 &Data& &game-45s& & &dance-45s& & &inter-45s& & &vlog-45s& \\
Model&Scale&x2&x3&x4&x2&x3&x4&x2&x3&x4&x2&x3&x4\\
\hline
 &M0&35.42&30.63&28.65&43.12&36.62&34.95&38.64&31.97&28.32&45.71&41.40&39.20\\
ESPCN&$S_{1-n}$&35.55&30.67&28.74&43.27&36.72&35.09&38.81&32.14&28.61&45.81&41.52&39.29
\\
 &Ours&36.09&31.06&29.05&43.56&36.89&35.30&38.88&32.22&28.75&46.19&41.72&39.52\\
\hline
 &M0&35.05&30.50&28.59&42.67&36.79&34.60&38.66&31.78&28.25&45.87&41.58&39.29\\
SRCNN&$S_{1-n}$&35.15&30.55&28.61&42.69&36.90&34.69&38.79&31.93&28.38&45.95&41.66&39.36\\
 &Ours&35.49&30.63&28.66&43.06&37.01&34.86&38.88&32.02&28.48&46.18&41.85&39.52\\
\hline
 &M0&40.29&34.53&31.28&45.03&37.95&36.57&41.99&33.80&30.34&47.61&42.92&40.94
\\
VDSR&$S_{1-n}$&41.37&34.92&32.42&45.18&38.05&36.82&42.40&34.53&31.10&47.88&43.33&41.23
\\
 &Ours&41.92&35.56&33.16&45.41&38.24&37.16&42.86&34.49&30.95&48.00&43.50&41.38\\
\hline
 &M0&42.11&35.75&33.33&45.49&38.26&37.32&42.73&34.49&31.34&47.98&43.58&41.53
\\
EDSR&$S_{1-n}$&43.22&36.72&34.32&45.71&38.61&37.67&43.31&35.80&32.67&48.48&44.12&42.12
\\
 &Ours&43.32&37.19&34.61&46.00&39.96&37.84&43.37&35.62&32.35&48.45&44.11&42.16\\
\hline

\end{tabular}
\end{center}
\caption{The generalization ability of our method on various SR architecture.}
\label{tab:gen}
\end{table*}
In this section, we report the results on VSD4K. We mainly present the results of 15s, 30s, 45s, 1min, 2min, and 5min across scaling factors $\times 2$,$\times 3$, and $\times 4$. We compare our results with two other methods. The first method trains a content-aware DNN for the whole video, which is denoted as M0. The second method divides a video into $n$ chunks and trains one DNN for each chunk \cite{yeo2018neural}. We denote the second method as ${S_{1 - n}}$. Compared with ${S_{1 - n}}$ \cite{yeo2018neural}, our methods significantly reduce the parameters of models. As shown in Tab. \ref{tab:vsd4k}, our methods can consistently obtain slightly better performance with less parameters compared with ${S_{1 - n}}$ \cite{yeo2018neural}. We also show the qualitative comparison in Fig. \ref{fig:vsd4k}.
\subsection{Ablation study}
\label{sec:abl}
{\bf Variants of CaFM}
Since we implement CaFM as $1 \times 1$ depth-wise convolution, we also study the effect of different kernel sizes for depth-wise convolution. We compare the results of $1 \times 1$, $3 \times 3$, $5 \times 5$, and $7 \times 7$ as shown in Tab. \ref{tab:abl1}. We also show the percentage of CaFM's parameter compared with EDSR in Tab. \ref{tab:abl1}. In general, larger kernel can achieve slightly better performance but will result in more parameters. Nevertheless, CaFM with $1 \times 1$ kernel already achieves competitive performance. 

\begin{table}[t]
	\begin{center}
		\setlength{\belowcaptionskip}{2.7pt}
		\centering
		\begin{tabular}{p{0.1\textwidth}p{0.08\textwidth}<{\centering}|p{0.04\textwidth}<{\centering}p{0.04\textwidth}<{\centering}p{0.04\textwidth}<{\centering}p{0.04\textwidth}<{\centering}}
			\hline
			\textbf{Dataset:} & game-15s & 1x1 & 3x3 & 5x5 & 7x7  \\
			\hline
			\textbf{PSNR:} & x2 & 43.13 & 43.16 & 43.17 & 43.23 \\
			& x3 & 37.04 & 37.09 & 37.10 & 37.18 \\
			   & x4 & 34.47 & 34.45 & 34.50 & 34.54 \\
			\hline
			\textbf{Percentage:} & x2 & 0.66\% & 2.20\% & 4.62\% & 8.19\% \\
			& x3 & 0.58\% & 1.96\% & 3.74\% & 7.23\% \\
			  & x4 & 0.60\% & 1.67\% & 3.83\% & 7.40\% \\
			\hline
		\end{tabular}
	\end{center}
	\caption{The above part shows the PSNR of different kernel size in 15s game video. The below part demonstrates the parameter percentage of CaFM compared with EDSR. }
	\label{tab:abl1}
	\vspace{-0.4cm}
\end{table}

{\bf Benefit of Joint Training}
As described above, joint training is important for our method to achieve good performance. To evaluate the benefit of joint training, we first train the DNN on a specific video from scratch, denoting $M_0$. Then we freeze the parameters and add CaFM for each chunk. We finetune the parameters of CaFM to overfit each chunk. This result is denoted as FT in Tab. \ref{tab:abl2}. The results of training separate DNN for each chunk are also reported as ${S_{1 - n}}$ \cite{yeo2018neural}. Our results outperform the results of other methods, demonstrating the benefit of joint training.

\begin{table}[t]
	\begin{center}
		\setlength{\belowcaptionskip}{2.7pt}
		\centering
		\begin{tabular}{p{0.09\textwidth}p{0.03\textwidth}|p{0.06\textwidth}<{\centering}|p{0.06\textwidth}<{\centering}|p{0.06\textwidth}<{\centering}|p{0.04\textwidth}<{\centering}}
			\hline
			  &  & M0 & $S_{1-n}$ & FT & Ours  \\
			\hline
			  &x2&42.24&42.82&42.30&43.13 \\
			game-15s&x3&35.88&36.42&35.93&37.04\\
			  &x4&33.44&34.00&33.49&34.47\\
			\hline
			 &x2&44.35&44.48&44.37&44.71\\
			dance-15s&x3&37.57&37.69&37.59&37.82\\
			 &x4&36.18&36.40&36.22&36.61\\
			\hline
		\end{tabular}

	\end{center}
	\caption{Benefit of joint training. }
	\label{tab:abl2}
	\vspace{-0.4cm}
\end{table}

\subsection{The generalization of our method}
\label{sec:gen}
As shown in Tab. \ref{tab:gen}, we present the results of our method applied to different SR backbones. We select four 45s videos and adopt 4 popular SR backbones: SRCNN \cite{dong2014learning}, ESPCN \cite{shi2016real}, VDSR \cite{kim2016accurate} and EDSR \cite{lim2017enhanced}. These experiments aim to study the generalization ability of our method. The conclusion is consistent with the results shown above. The joint training framework along with CaFM can generalize well to other SR architectures, validating the generalization ability of our method.
\subsection{Comparisons with H.264/H.265}
\label{sec:h264}
Our method can be also viewed as a video coding approach. Therefore, we conduct primary experiments to compare our method with commercial H.264 and H.265 standard. For H.264 and H.265, we decrease the bitrate of the video while maintaining the resolution to obtain a video with the same size as our method (LR video and models). We compare our SR video with low bitrate videos of H.264 and H.265. Six videos are randomly selected from VSD4K to conduct the comparison with H.264 and H.265. The quantitative results are shown in Tab. \ref{tab:h264}. Our results outperform H.264 and H.265 in most cases. We also show the qualitative comparison in Fig. \ref{fig:vsd4k}.\\
This primary experiment shows the great potential of our method. Furthermore, since the proposed CaFM takes less than $1\%$ additional parameters for each chunk, we believe our method can achieve better results by dividing the video into more chunks. However, this will result in more training time accordingly.
\section{Future Work}
Although our method utilizes the overfitting property of DNN to achieve significant SR performance, it needs to train one network for each video chunk. This brings additional cost for network training. In order to further improve the video delivery systems, we believe reducing the training time is a promising and important future work. Learning methods like MAML \cite{finn2017model} can be used to accelerate the training process of SR networks. Recently, MAML has been applied to zero-shot image SR \cite{park2020fast,soh2020meta}, aiming to reduce the computational cost of DNN training.

\section{Conclusion}
In this paper, we study the problem of applying DNN to video delivery. Neural video delivery adopts the overfitting property of DNN, by training one model for each video chunk. In order to avoid streaming one model for each video chunk, we propose a novel joint training framework along with the CaFM module. With our method, each video chunk only requires less than $1\% $ of original parameters to be streamed and achieve better SR performance. We conduct comprehensive analysis of our method to demonstrate its advantages. Besides, we also compare with the commercial H.264 and H.265 standards, showing the potential of our approach. We hope our work can inspire future work on applying DNN to video delivery.


\begin{figure*}[!t]
\begin{center}
\includegraphics[width=15cm, height=12cm]{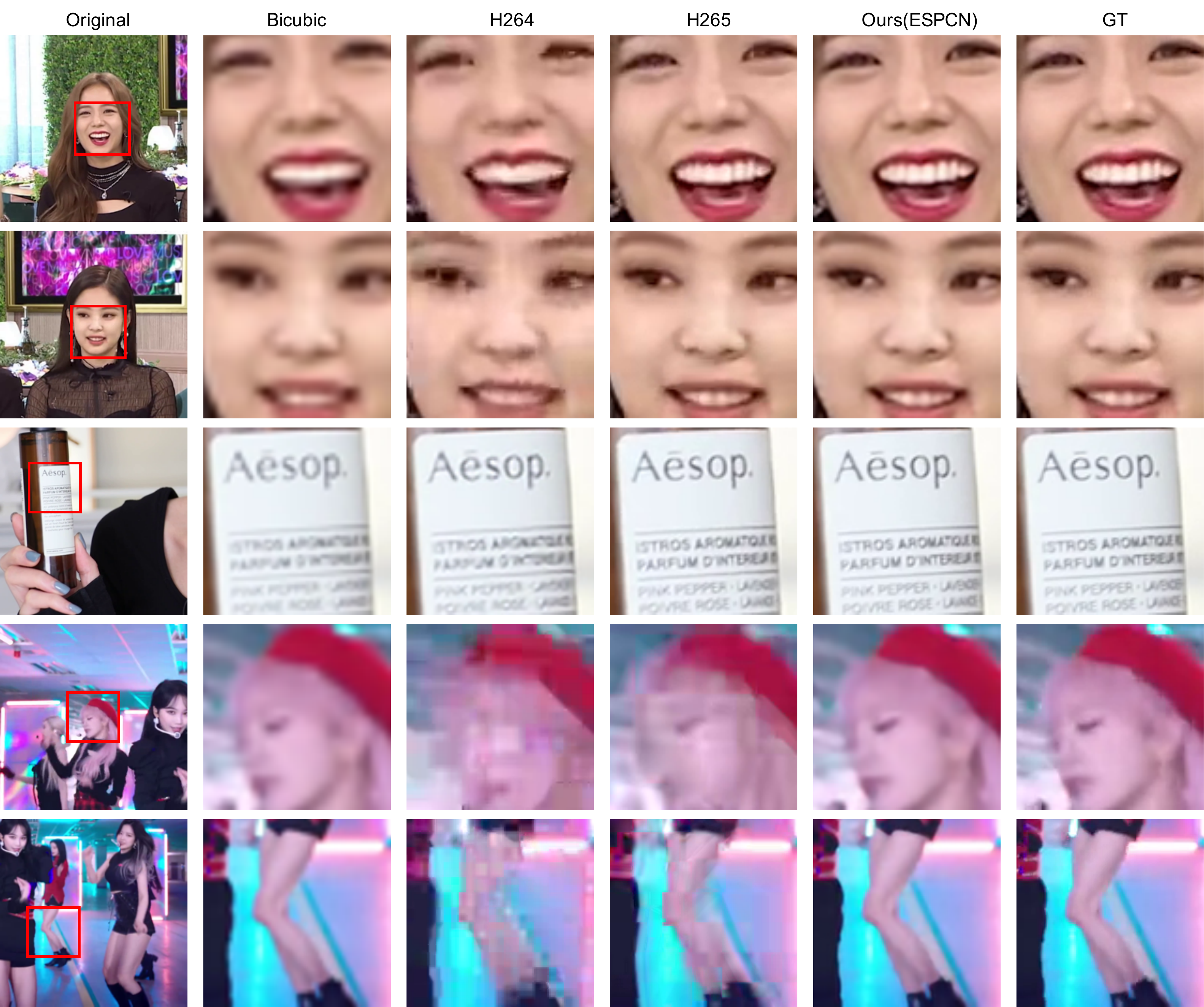}
\end{center}
   \caption{Qualitative comparisons with H.264/H.265. We use a lightweight model (ESPCN) in these comparisons. Best viewed by zooming x4.}
\label{fig:1}
\end{figure*}

\begin{table*}[!t]
\scriptsize
\begin{center}
\begin{tabular}{|p{0.10\textwidth}<{\centering}|p{0.18\textwidth}<{\centering}|p{0.12\textwidth}<{\centering}|p{0.12\textwidth}<{\centering}|p{0.10\textwidth}<{\centering}|p{0.06\textwidth}<{\centering}|p{0.08\textwidth}<{\centering}|}
\hline
Category&Source&Highest Resolution&Training Resolution&Bit-rate (Mbit/s)&FPS&Video Length\\
\hline\hline
Game & LoL Game: \url{https://www.youtube.com/watch?v=BQG92HATfvE} & $3840\times2160$ &$1920\times1080$&10.04 &30 &15s-5min\\\hline
Vlog & Make-up tutorial: \url{https://www.youtube.com/watch?v=MYGZ2_X5L3E} & $3840\times2160$ &$1920\times1080$&10.10 &30&15s-5min\\\hline
Inter & Blackpink interview: \url{https://www.youtube.com/watch?v=6FBCVpU3XG4}  & $3840\times2160$ &$1920\times1080$&10.15 &30&15s-5min\\\hline
Sport & Extreme sports: \url{https://www.youtube.com/watch?v=M0jmSsQ5ptw} & $3840\times2160$ &$1920\times1080$&10.04 &30&15s-5min\\\hline
Dance & izone performance: \url{https://www.youtube.com/watch?v=hBlLaEt1VjI}  & $3840\times2160$ &$1920\times1080$&10.03 &30&15s-5min\\\hline
City & London city drive: \url{https://www.youtube.com/watch?v=QI4_dGvZ5yE}  & $3840\times2160$ &$1920\times1080$&9.91 &30&15s-5min\\\hline
\end{tabular}
\end{center}
\caption{Details of VSD4K datasets.}
\label{tab:content}
\end{table*}

\begin{table*}[!t]
\scriptsize
\begin{center}
\begin{tabular}{|p{0.09\textwidth}<{\centering}|p{0.03\textwidth}<{\centering}p{0.07\textwidth}<{\centering}p{0.03\textwidth}<{\centering}|p{0.03\textwidth}<{\centering}p{0.07\textwidth}<{\centering}p{0.03\textwidth}<{\centering}|p{0.03\textwidth}<{\centering}p{0.06\textwidth}<{\centering}p{0.03\textwidth}<{\centering}|p{0.03\textwidth}<{\centering}p{0.07\textwidth}<{\centering}p{0.03\textwidth}<{\centering}|}

\hline
 & &game-45s& & &dance-45s& & &inter-45s& & &vlog-45s& \\\hline
Method&x2&x3&x4&x2&x3&x4&x2&x3&x4&x2&x3&x4\\
\hline
H.264&\textcolor{blue}{37.72}&\textcolor{blue}{33.43}&\textcolor{blue}{30.63}&29.12&24.51&21.86&36.54&\textcolor{blue}{33.26}&\textcolor{blue}{31.02}&42.44&39.79&37.65\\\hline
H.265&\textcolor{red}{38.32}&\textcolor{red}{34.56}&\textcolor{red}{32.28}&\textcolor{blue}{30.90}&\textcolor{blue}{27.09}&\textcolor{blue}{24.86}&\textcolor{blue}{36.94}&\textcolor{red}{33.92}&\textcolor{red}{31.85}&\textcolor{blue}{43.39}&\textcolor{blue}{41.04}&\textcolor{blue}{39.13}\\\hline
Ours(ESPCN)&36.09&31.06&29.05&\textcolor{red}{43.56}&\textcolor{red}{36.89}&\textcolor{red}{35.30}&\textcolor{red}{38.88}&32.22&28.75&\textcolor{red}{46.19}&\textcolor{red}{41.72}&\textcolor{red}{39.52}\\\hline
Storage(MB)&14.46&6.48&3.90&14.08&6.39&3.80&13.97&6.38&3.79&14.00&6.37&3.78\\\hline
\end{tabular}
\end{center}
\caption{Quantitative comparisons with H.264/H.265. We use a lightweight model (ESPCN) in these comparisons. Red and blue indicate the best and the second best results.}
\label{tab:h264-2}
\end{table*}

\begin{table*}[!th]
\scriptsize
\begin{center}
\begin{tabular}{|p{0.14\textwidth}<{\centering}|p{0.1\textwidth}<{\centering}|p{0.07\textwidth}<{\centering}|p{0.05\textwidth}<{\centering}p{0.03\textwidth}<{\centering}p{0.05\textwidth}<{\centering}p{0.03\textwidth}<{\centering}|p{0.04\textwidth}<{\centering}|p{0.03\textwidth}<{\centering}p{0.03\textwidth}<{\centering}p{0.03\textwidth}<{\centering}p{0.03\textwidth}<{\centering}|p{0.04\textwidth}<{\centering}|}

\hline
 & & & & &Vid4& & \\
Method&Model&Dataset&Calender&City&Foliage&Walk&Average\\\hline
External learning & EDVR-M\cite{wang2019edvr}&REDS&21.82&25.91&24.67&28.83&25.31\\
 & EDVR-L\cite{wang2019edvr}&REDS&21.89&25.68&24.77&29.17&25.38\\
  & EDVR-L\cite{wang2019edvr}&Vimeo-90K&22.18&26.30&25.00&29.55&25.76\\
\hline
Content-aware learning & EDSR-M&Vid4&25.23&30.56&26.48&31.00&28.32\\
\hline
Content-aware learning & EDSR-L&Vid4&\textcolor{red}{27.19}&\textcolor{red}{32.19}&\textcolor{red}{27.66}&\textcolor{red}{32.47}&\textcolor{red}{29.88}\\
\hline\hline
 & & & & &REDS& & \\
Method&Model&Dataset&000&011&015&020&Average\\\hline
External learning & EDVR-M\cite{wang2019edvr}&REDS&27.72&31.26&33.42&29.57&30.49\\
 & EDVR-L\cite{wang2019edvr}&REDS&\textcolor{red}{28.01}&32.17&34.06&\textcolor{red}{30.09}&31.09\\
  & EDVR-L\cite{wang2019edvr}&Vimeo-90K&27.80&31.03&33.45&29.50&30.45\\
\hline
Content-aware learning & EDSR-M&REDS&27.27&31.31&34.02&29.07&30.42\\
\hline
Content-aware learning & EDSR-L&REDS&27.63&\textcolor{red}{32.38}&\textcolor{red}{34.94}&29.86&\textcolor{red}{31.20}\\\hline
\end{tabular}
\end{center}
\caption{Comparisons of content-aware learning versus external learning. EDVR-M, EDVR-L, EDSR-M, EDSR-L has 10, 40, 16, 32 resblocks respectively. Red indicates the best results.}
\label{tab:2}
\end{table*}

\begin{table*}[!th]
\scriptsize
\begin{center}
\begin{tabular}{|p{0.09\textwidth}<{\centering}|p{0.03\textwidth}<{\centering}p{0.07\textwidth}<{\centering}p{0.03\textwidth}<{\centering}|p{0.03\textwidth}<{\centering}p{0.07\textwidth}<{\centering}p{0.03\textwidth}<{\centering}|}

\hline
 & &90027457& & &72549854&  \\\hline
Method&x2&x3&x4&x2&x3&x4\\
\hline
M0                & 49.86    & 45.60     & 43.55    & 40.22    & 35.21    & 32.50    \\
			$S_{1-n}$              & 50.01    & 46.00       & 44.07    & 40.32    & 35.47    & 32.84   \\
			Ours        & \textcolor{red}{50.14}  & \textcolor{red}{46.33}   & \textcolor{red}{44.15}    & \textcolor{red}{40.41}    & \textcolor{red}{35.40}     & \textcolor{red}{32.73}  \\
			Margin            & +0.13    & +0.33    & +0.08     & +0.09     & -0.07    & -0.11
			\\\hline
			H.264&41.84&40.33&39.18&33.10&32.05&31.06\\
			H.265&\textcolor{blue}{42.02}&\textcolor{blue}{40.81}&\textcolor{blue}{39.29}&\textcolor{blue}{33.22}&\textcolor{blue}{32.55}&\textcolor{blue}{31.95}\\\hline
			Size(MB)&24.10&14.41&10.97&13.45&10.02&8.38\\\hline
\end{tabular}
\end{center}
\caption{PSNR results on public Vimeo-90K dataset. Red and blue indicate the best and the second best results among our method, H.264, and H.265.}
\label{tab:vimeo}
\end{table*}

\section{Appendix}
In Sec. \ref{sec:data}, we demonstrate the details of our VSD4K dataset. As reported in Sec. \ref{sec:content}, we also evaluate content-aware learning and external learning on public datasets like Vid4 \cite{liu2013bayesian} and REDS \cite{nah2019ntire}. In Sec. \ref{sec:264}, we apply our method to lightweight architecture (ESPCN \cite{shi2016real}) and compare with H.264/H.265 standard. We evaluate our method on public vimeo dataset \cite{xue2019video} in Sec. \ref{sec:vimeo}. All the results use PSNR as the evaluation metric.
\subsection{Details of VSD4K}
\label{sec:data}
As shown in Tab. \ref{tab:content}, we download the original 4K videos from YouTube as our source videos. Due to computational limitation, we resize the source videos to 1080p as our ground-truth. According to FFmpeg \cite{ffmpeg}, we resize the 4k video by bicubic interpolation and alternate bit-rate based on \cite{TP-toolbox-web}.

\subsection{Content-aware learning on public datasets}
\label{sec:content}
We present the benefit of utilizing DNN's overfitting property for video delivery on public dataset. As shown in Tab. \ref{tab:2}, we compare content-aware learning and external learning on public datasets like Vid4 \cite{liu2013bayesian} and REDS \cite{nah2019ntire}. As can be seen, EDSR with content-aware learning significantly outperforms EDVR with external learning. These results prove that content-aware learning is more suitable for video delivery compared with external learning.

\subsection{H.264/H.265 against Ours (ESPCN)}
\label{sec:264}
In this section, we adopt ESPCN \cite{shi2016real} to compare our method with H.264/H.265 standard under same storage cost. The quantitative results are shown in Tab. \ref{tab:h264-2}. Our results still outperform H.264 and H.265 in most cases. We also show the qualitative comparison in Fig. \ref{fig:1}.

\subsection{Evaluation on Vimeo90k\cite{xue2019video}}
\label{sec:vimeo}
In this section, we conduct experiments on public Vimeo90k\cite{xue2019video} to present the universality of our method. We randomly select two videos from \url{http://data.csail.mit.edu/tofu/dataset/original_video_list.txt}. As shown in Tab. \ref{tab:vimeo}, our method outperforms $S_{1-n}$ to some extent. We also compare our method with standard H.264 and H.265. For a particular LR video, we set the sum of (LR video and SR model) as constant value. Then, we decrease the bit-rate of H.264 and H.265 video to reach the same storage as the former. Under some storage cost, our method shows promising results.

\clearpage
{\small
\bibliographystyle{ieee_fullname}
\bibliography{egbib}
}

\end{document}